\documentclass[aps,prb,amsmath,twocolumn,amssymb,floatfix,showpacs,superscriptaddress,10pt]{revtex4-2}
\usepackage{placeins}
\pdfoutput=1
\usepackage[dvips]{graphics}
\usepackage{bm}
\usepackage{epsfig}
\usepackage{enumerate}
\usepackage{subfigure}
\usepackage[dvipsnames]{xcolor}
\usepackage{amsmath}
\usepackage{color}
\usepackage{braket}
\usepackage{graphicx}
\usepackage{float}
\usepackage{gensymb}
 \usepackage[utf8]{inputenc}
 \usepackage{array}
 \usepackage{tipa}
 \usepackage{graphicx}
\usepackage{tikz-cd}
\usepackage[normalem]{ulem}
\usepackage[colorlinks=true,linktoc=page,linkcolor=Blue,citecolor=Blue,urlcolor=Blue]{hyperref}
\usepackage{cleveref}
\usepackage[symbol]{footmisc}

\newcommand\bea{\begin{eqnarray}}
\newcommand\eea{\end{eqnarray}}
\newcommand\beq{\begin{equation}}  
\newcommand\eeq{\end{equation}}

\begin{document}
\title{Emergent Quasiparticles \& Field-Tuned RIXS Spectra in a Trimerized Spin-1/2 Chain}
\author{Subhajyoti Pal}
\email{subhajyoti.pal@niser.ac.in}
\author{Pradeep Thakur}
\author{Ashis Kumar Nandy}
\author{Anamitra Mukherjee}
\email{anamitra@niser.ac.in}
\affiliation{School of Physical Sciences, National Institute of Science Education and Research, Jatni 752050, India}
\affiliation{Homi Bhabha National Institute, Training School Complex, Anushaktinagar, Mumbai 400094, India}

\begin{abstract}
We investigate spin-flip excitations in the spin-1/2 trimer chain $\rm{Cu_3(P_2O_6OH)_2}$, featuring an antiferromagnetic exchange motif $J_1$–$J_1$–$J_2$ with $J_1 < J_2$. Using density matrix renormalization group (DMRG) simulations, we demonstrate that single-spin-flip processes induced by resonant inelastic X-ray scattering (RIXS) generate emergent gapless modes governed by the underlying trimer periodicity alongside distinct high-energy excitations. By combining exact diagonalization and real-space renormalization group (RG) techniques, we attribute these features to fractionalized spinons and composite quasiparticles arising from one- and two-trimer excitations. Furthermore, we show that multi-spin RIXS excitations yield experimentally distinguishable spectral signatures of composite modes absent in single-spin-flip spectra.
At the field-induced 1/3 magnetization plateau, single-spin-flip RIXS spectra evolves with the magnetic field to favor spin-polarized composite quasiparticles. This trend culminates in a gapless spectrum of spin-1 excitations beyond the plateau, paving the way for field-tuned Bose condensation of composite modes.
\end{abstract}

\maketitle
\textit{Introduction:---}  
Low-dimensional quantum magnets offer a rich platform for realizing exotic quantum phases and fractionalized excitations. In particular, spin-1/2 chains and ladders, shaped by strong correlations and reduced dimensionality, have enabled deep insights into confinement–deconfinement transitions~\cite{frac-qpt,deconf-2d-prx,deconf-2d-prb}, quantum criticality~\cite{1d-qcp-1,1d-qcp-2,1d-qcp-3,1d-qcp-4,1d-qcp-5,1d-qcp-6}, and spin fractionalization~\cite{frac-haldane,frac-1,frac-2,frac-3,frac-kondo}. Spin systems with complex unit cells further enrich one-dimensional (1D) physics by inducing novel quasiparticles and emergent transitions~\cite{unit_cell-1,bera2022emergent,cheng2022fractional,prabhakar2024}. These advances are fueled by the synthesis of model (1D) compounds~\cite{1dcuprate1,PhysRevLett.77.4054,PhysRevB.45.5744} and modern spectroscopies such as inelastic neutron scattering (INS), resonant inelastic X-ray scattering (RIXS)~\cite{rixs-rev-nat,rixs-prx-1,rixs-prx-2,RevModPhys.83.705}, spin echo~\cite{spin-echo-spinon}, and terahertz probes~\cite{thz-frac}.

The spin-1/2 Heisenberg antiferromagnetic chain (HAC) epitomizes integrable 1D systems with gapless spinon excitations~\cite{Bethe1931}. However, integrability is lost in systems with exchange modulation or frustration. A prominent example is $\rm{Na_2Cu_3Ge_4O_{12}}$, a trimer chain with $J_1$–$J_1$–$J_2$ exchange ($J_1 > J_2$), where INS has revealed both gapless spinons and gapped composite modes—doublons and quartons—arising from trimer-localized excitations dressed by spinons~\cite{bera2022emergent,prabhakar2024}.

\begin{figure}[t]
  \centering
      \includegraphics[width=0.98\linewidth]{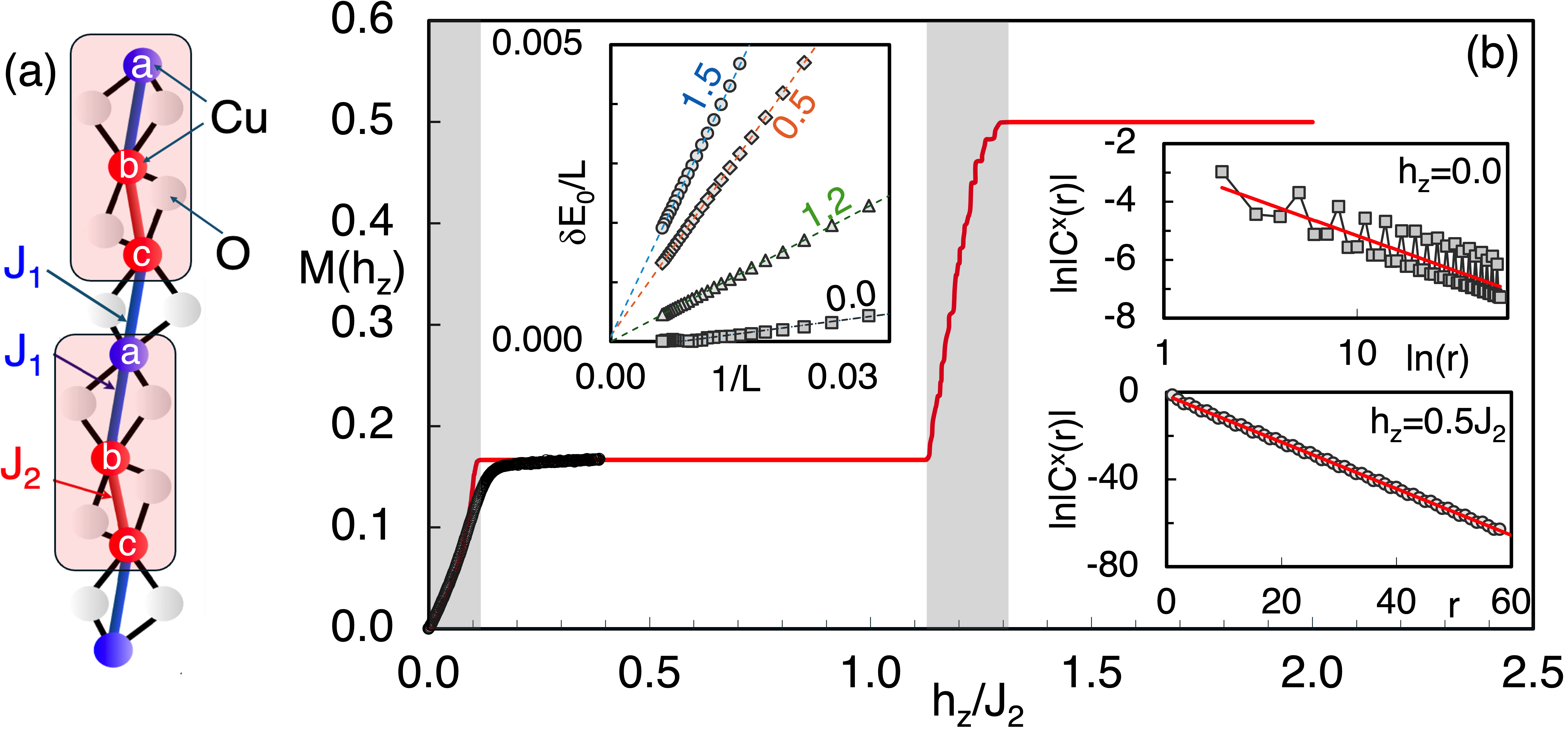}
  \caption{\textbf{Spin trimer \& magnetization:} (a) Trimer structure of $\rm{Cu_3(P_2O_6OH)_2}$ with unit cells (rectangles). (b) Magnetization $M(h_z)$ (solid line) showing a 1/3 plateau ($M = 1/6$) and saturation ($M = 0.5$), with experimental data~\cite{hase20061} (open circles). Insets: left—finite-size scaling of the gap; right—$C^x(r) = \langle S^x_{L/2} S^x_{L/2 + r} \rangle$ showing power-law (top) vs. exponential decay (bottom) for gapless and gapped phases respectively.}
  \label{fig0}
\end{figure}

In contrast, the unfrustrated $\rm{Cu_3(P_2O_6OH)_2}$ [Fig.~\ref{fig0}(a)] hosts the reverse hierarchy ($J_1 < J_2$)~\cite{hase20061}, eliminating the weakly coupled trimer scenario valid for $\rm{Na_2Cu_3Ge_4O_{12}}$. This raises a fundamental question: how do spin-flip excitations evolve across the Heisenberg point ($J_1 = J_2$)~\cite{cheng2022fractional}, and are the excitation spectra qualitatively altered in the $J_1 < J_2$ regime? Notably, this material exhibits a robust 1/3 magnetization plateau~\cite{hase20061}, suggesting emergent periodicity, yet the nature and evolution of excitations within the plateau remain unexplored. Also, prior INS studies have primarily accessed single spin-flip modes; the role of nonlocal, multi-spin RIXS channels in generating novel excitations remains unchartered. To address these questions, we investigate spin-flip excitations in $\rm{Cu_3(P_2O_6OH)_2}$ using a combination of density matrix renormalization group (DMRG), exact diagonalization (ED), and effective low-energy modeling. Our analysis covers both single-spin inelastic neutron scattering (INS) and multi-spin resonant inelastic X-ray scattering (RIXS) processes at zero field and across the 1/3 magnetization plateau.

\begin{figure*}[t]
  \centering
      \includegraphics[width=1\linewidth]{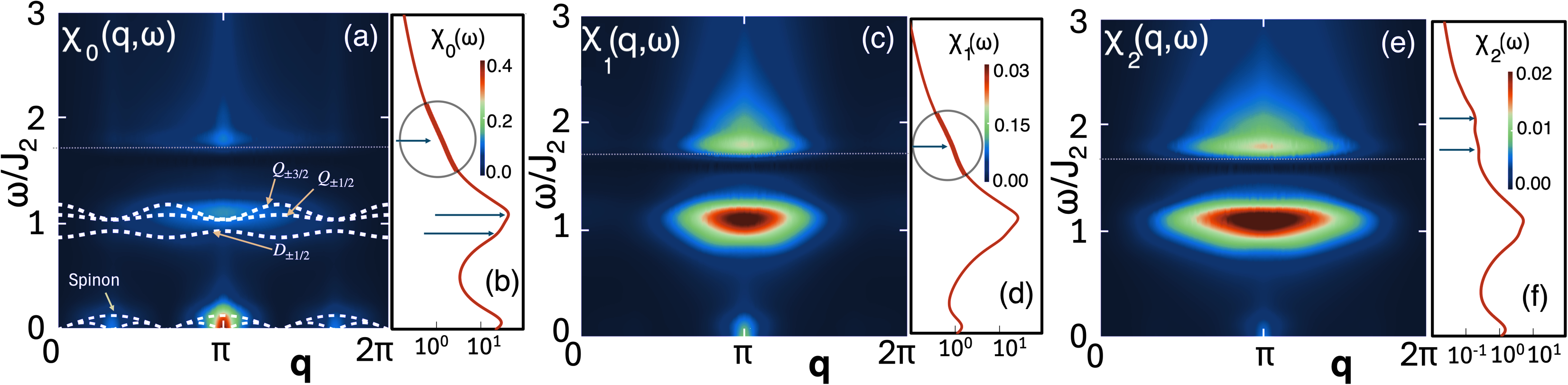}
  \caption{\textbf{Zero-field RIXS spectra:} (a,c,e) DMRG results for $\chi_l(\mathbf{q},\omega)$ with $l=0,1,2$ on 120 sites. (b,d,f) Corresponding $\mathbf{q}$-integrated ED spectra on 15 sites. In (a), the dashed curve marks the spinon continuum; solid lines show doublon and quarton dispersions. Data above $\omega = 1.8J_2$ is scaled by $100$ in all panels.}
  \label{fig1}
\end{figure*}
We find that the conventional spinon continuum of the Heisenberg antiferromagnetic chain is significantly reconstructed: a gapless low-energy mode emerges with momentum signatures reflecting the underlying trimer periodicity, accompanied by distinct high-energy structures. An effective spinon description captures the low-energy sector, while ED reveals that one- and two-trimer excitations form the basis of the higher-energy composite quasiparticles. Crucially, nonlocal multi-spin RIXS processes selectively enhance these composite features while suppressing spinon-like responses. As the magnetic field is tuned across the 1/3 magnetization plateau, the single spin-flip RIXS features evolve linearly with the field strength. Beyond the plateau, we uncover a gapless continuum of S=1 composite excitations, bridging the gapped trimer plateau and the fully polarized phase—signaling a possible crossover regime governed by Bose condensed composite quasiparticles.

\textit{Hamiltonian \& RIXS perturbation:---}  
We model $\rm{Cu_3(P_2O_6OH)_2}$ as a spin-1/2 antiferromagnetic trimer chain with $J_1$–$J_1$–$J_2$ couplings [Fig.~\ref{fig0}(a)], with ($J_1/J_2 = 0.27$) consistent with structural data~\cite{supp} (Sec. I), by the following Hamiltonian :
\begin{equation}
H = \sum_i \left(J_1\, {\bf S}_i^a \cdot {\bf S}_i^b + J_2\, {\bf S}_i^b \cdot {\bf S}_i^c + J_1\, {\bf S}_i^c \cdot {\bf S}_{i+1}^a \right),
\end{equation}
where $i$ indexes unit cells, and $\{a,b,c\}$ label trimer sublattices.

RIXS is a photon-in, photon-out probe described by the Kramers–Heisenberg formalism: an incident X-ray excites a $2p$ core electron to the $3d$ shell, creating a core-hole that decays radiatively, leaving behind valence excitations. At the Cu $L$-edge, strong $2p$ spin-orbit coupling permits spin-flip transitions ($\Delta S^z = \pm1$), defining the non-spin-conserving (NSC) channel. Within the ultra-short core-hole lifetime (UCL) approximation~\cite{RevModPhys.83.705,PhysRevX.6.021020,UKumar2022,PhysRevB.108.214405,fast-coll}, the RIXS cross-section is expanded in inverse core-hole lifetime $\Gamma$ (see \cite{supp}, Sec. II). The $l$-th order intensity is:
\begin{align}
\chi^l(\mathbf{q}, \omega) =& \frac{1}{\Gamma^{2l+2}} \sum_f \left| \left\langle f \left| \frac{1}{\sqrt{L}} \sum_i e^{i \mathbf{q} \cdot \mathbf{R}_i} O_{i,l} \right| g \right\rangle \right|^2 \nonumber \\
& \times \delta(E_f - E_g - \omega),
\label{eq:RIXS_Int_f}
\end{align}
with $O_{i,0} = S_i^x$,  
$O_{i,1} = \sum_{j \in \text{NN}(i)} J_{ij} S_i^x\, \mathbf{S}_i \cdot \mathbf{S}_j$, and  
$O_{i,2} = \sum_{j \ne k \in \text{NN}(i)} J_{ij} J_{ik} S_i^x\, (\mathbf{S}_i \cdot \mathbf{S}_j)(\mathbf{S}_i \cdot \mathbf{S}_k)$.  
Here, $|g\rangle$ and $|f\rangle$ are the ground and excited states of $H$, with energies $E_g$ and $E_f$, respectively. The $l=0$ term corresponds to single-spin-flip excitations, while higher orders ($l > 0$) encode multi-spin correlations.

\textit{Fractional magnetization plateau:---}  
We compute the ground state and magnetization using MPS-DMRG (iTensor~\cite{itensor}) for $L = 120$. Magnetization plateaus follow $N_0(S - M) = I$, where $N_0$ is spins per unit cell, $S$ the spin magnitude, $M$ the magnetization per site, and $I$ an integer. For $I=1$, this yields $M = 1/6$, or a 1/3 plateau~\cite{PhysRevLett.78.1984}. The magnetization $M(h)$ [Fig.~\ref{fig0}(b)] shows a 1/3 plateau between $0.179J_2$ and $1.1J_2$, and full polarization above $1.3J_2$. With $J_2 = 110$~K, this corresponds to $14$–$85$~$T$, matching experiments~\cite{hase20061,hase2007} (open circles), using $h_z(T) = h_z J_2 / (g \mu_B)$ with $g = 2.12$.
Shaded regions in Fig.~\ref{fig0}(b) mark gapless behavior, confirmed by finite-size scaling of the first excitation gap $\delta E_0/L$ (left inset), which extrapolates to zero, unlike for the plateau and polarized state. For the gapless phases, the spin correlations $C^x(r) = \langle S^x_{L/2} S^x_{L/2 + r} \rangle$ decay as $1/r$ at $h_z = 0$ [top right inset], with exponent $\eta = 1$, consistent with conformal charge $c = 1$ and $S=1/2$ HAC universality class~\cite{universality-1} (see \cite{supp}, Sec. III). In contrast, exponential decay of $C^x(r)$ in the plateau indicates a gapped phase~\cite{koma2007} [bottom right inset].

\begin{figure*}[t]
  \centering
      \includegraphics[width=0.9\linewidth]{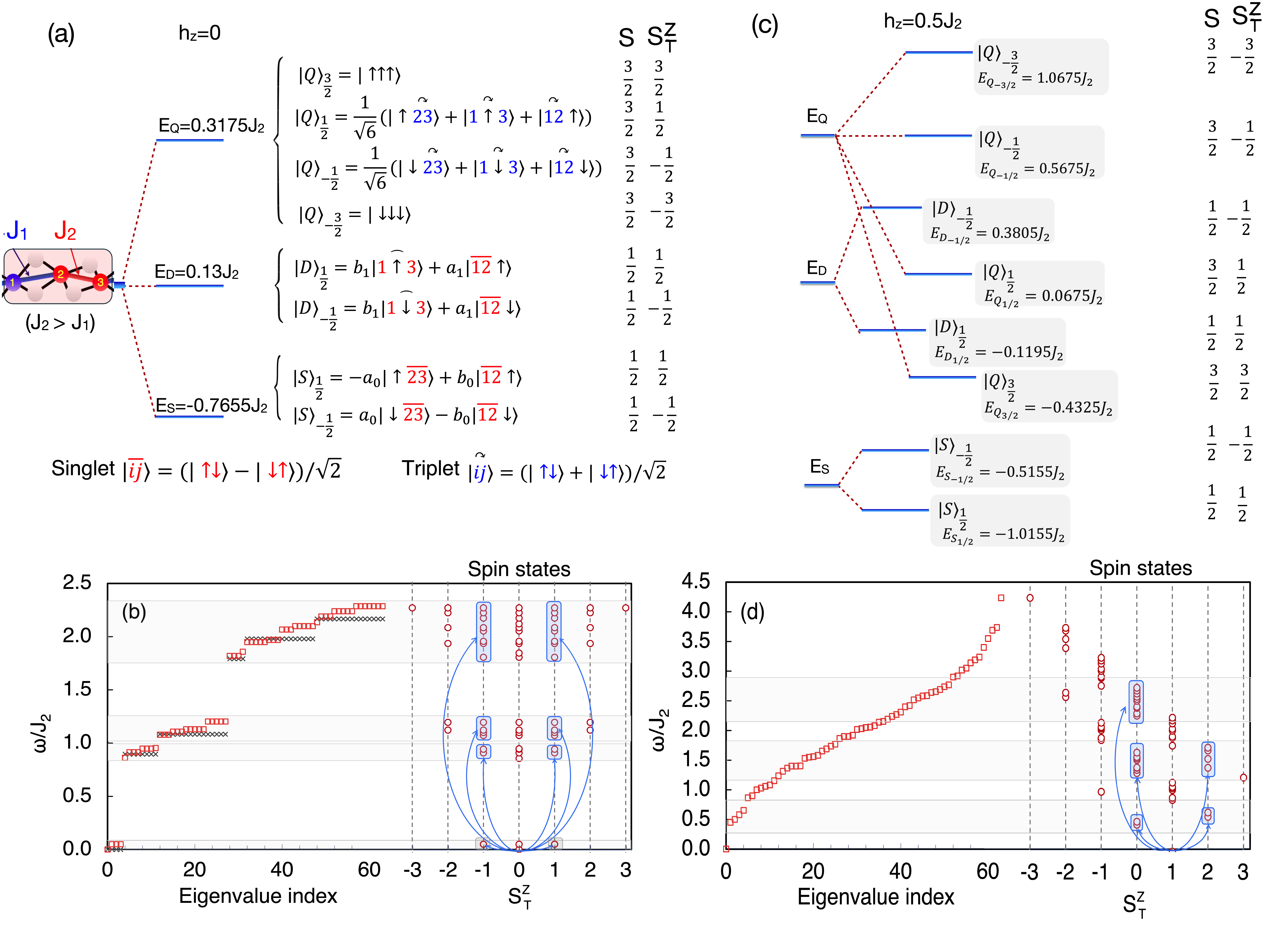}
  \caption{\textbf{Trimer spectra:} (a,c) Single-trimer eigenspectrum and spin configurations at $h_z = 0$ and $0.5J_2$, with site labels defined in the trimer schematic (a). Coefficients of the eigenstates are  $a_0 = 0.91531$, $b_0=0.151984$, $a_1=0.4407$ and $b_1=0.703949$. (b) Two-trimer spectra at $h_z = 0$ and $0.5J_2$: crosses show decoupled levels, squares show coupled levels, and $S^z_T$ labels are shown in the right. Blue arrows mark allowed $\Delta S^z = \pm 1$ transitions. (d) provides the same information for two coupled trimers for $h_z=0.5J_2$.}
  \label{fig2}
\end{figure*}

\textit{Zero-field RIXS:---}  
We analyze $\chi^{0}(\mathbf{q},\omega)$ at $l=0$ [Fig.~\ref{fig1}(a)], revealing three distinct features. The lowest is a gapless mode extending up to $\omega/J_2 \sim 0.135$, with weight concentrated at $\mathbf{q} = \frac{(2n+1)\pi}{3}$ ($n = 0,1,2$). A mid-energy feature appears at $\mathbf{q} = \pi$, $\omega = 1.11J_2$, with two close peaks in the $\mathbf{q}$-integrated spectra [Fig.~\ref{fig1}(b), arrows]. A weak high-energy bump occurs near $\omega \sim 1.8J_2$ (encircled arrow). 
We now use ED to resolve the structure of $\chi^0(\mathbf{q},\omega)$.

We begin with a single isolated trimer, which has three eigenstates [Fig.~\ref{fig2}(a)]: a doubly-degenerate $S=1/2$ ground state $|S\rangle_\sigma$ with energy $E_S=-0.7655J_2$, excited (doubly-degenerate) $|D\rangle_\sigma$ states at $E_D = 0.1305J_2$, and a fourfold $S=3/2$ multiplet $|Q\rangle_\sigma$ at $E_Q = 0.3175J_2$. Here, $\sigma$ denotes $S^z$ and are provided in Fig.~\ref{fig2}(a). The excitation energies $E_D - E_S = 0.896J_2$ and $E_Q - E_S = 1.083J_2$ match the mid-energy peaks in Fig.~\ref{fig1}(a) and (b) around $\omega=1.1J_2$, but cannot generate the high-energy features.

To understand the full spectra in a unified fashion, we study the two-trimer spectra [Fig.~\ref{fig2}(b)]. In the decoupled case, the Hilbert space of 64 states splits into sectors like $|SS\rangle$, $|SD\rangle$, $|SQ\rangle$, $|DD\rangle$, $|DQ\rangle$,  $|QQ\rangle$, etc., with a four-fold degenerate ground state belonging to $|SS\rangle$ with energy $2E_S$ and excitations $E_S + E_D$, $E_S + E_Q$, $2E_D$, $E_D + E_Q $, and $2E_Q$. Relative to the ground state energy, these occur at $0.896J_2$, $1.083J_2$, $1.792J_2$, $1.979J_2$, and $2.166J_2$, respectively (labeled by crosses in Fig.~\ref{fig2}(b)). Upon coupling, level repulsion broadens these into bands (gray regions), and the two-trimer ground state $|g\rangle_{2\rm Tr}$ becomes non-degenerate with dominant overlap with states in $|SS\rangle$ sector. Blue arrows on the right in Fig.~\ref{fig2}(b) mark allowed $\Delta S = \pm1$ transitions contributing to $\chi^{0}(\mathbf{q},\omega)$.

We first focus on the lowest band in Fig.~\ref{fig2}(b). Since $O_{i,0}=S^x_i$ causes local single spin flip, we investigate whether such single spin-flip fractionalizes—despite the trimerized unit cell. For this, we derive a low-energy effective model using a real-space RG approach\cite{rg-realspace,rg-delgado} (detailed in ~\cite{supp} (Sec. IV.A)):
\begin{equation}
\hat{H}_\text{eff}=J_\text{eff}\sum_{i} {\bf \tilde{S}}_i \cdot{\bf \tilde{S}}_{i+1} + J_\text{eff}'\sum_{i} {\bf \tilde{S}}_i \cdot{\bf \tilde{S}}_{i+2},
\end{equation}
with $J_\text{eff} = 0.16J_1$, $J_\text{eff}' \sim 10^{-4}J_1$. This yields a spin-1/2 Heisenberg chain with negligible NNN coupling. The analytic spinon continuum ranging from $\omega = J_\text{eff} |\sin(3\mathbf{q})|/2$ to $\omega = \pi J_\text{eff} |\sin(3\mathbf{q}/2)|$ (with $\mathbf{q} = (2n+1)\pi/{3}$), overlaid in Fig.~\ref{fig1}(a), matches the DMRG data—identifying the lowest feature as deconfined spinons.

The spin-allowed transitions (blue arrows) reveal two classes of high-energy excitations. The first involves single-trimer modes: $|SS\rangle$ components of the ground state are excited to $|SD\rangle$ or $|SQ\rangle$ sectors via a single spin flip. For instance, $|D\rangle_{\pm1/2}$ ($0.896J_2$) and $|Q\rangle_{\pm1/2,\pm3/2}$ ($1.083J_2$) match mid-energy peaks. Their dispersion over a staggered product ansatz $\prod_i^{L/3} |S\rangle_{(-1)^i 1/2}$ (see~\cite{supp}, Sec. IV.B) reproduces the centroids of the gray bands in Fig.~\ref{fig2}(b), as shown by dashed lines in Fig.~\ref{fig1}(a). These delocalized, spinon-dressed single-trimer excitations correspond to doublons and quartons~\cite{cheng2022fractional}. The second class involves true two-trimer composites and cannot be explained as excitations of a single-trimer: single spin flips connect $|SD\rangle$ and $|SQ\rangle$ sectors contributing to the ground state to $|DD\rangle$, $|DQ\rangle$, or $|QQ\rangle$. For instance, $|D\rangle_{-1/2}|D\rangle_{-1/2}$ at $1.791J_2$ underlies the $\omega \sim 1.8J_2$ feature, while $|Q\rangle_{3/2}|Q\rangle_{-1/2}$ and $|Q\rangle_{-1/2}|Q\rangle_{-1/2}$ at $2.08J_2$ define the lower and upper edges of highest gray band in Fig.~\ref{fig2}(b). Due to the dominance of the $|SS\rangle$ in the ground state, the corresponding features in Fig.~\ref{fig1}(a) are highly suppressed. This completes the characterization of all $l = 0$ excitations.
\begin{figure}[t]
  \centering
      \includegraphics[width=1\linewidth]{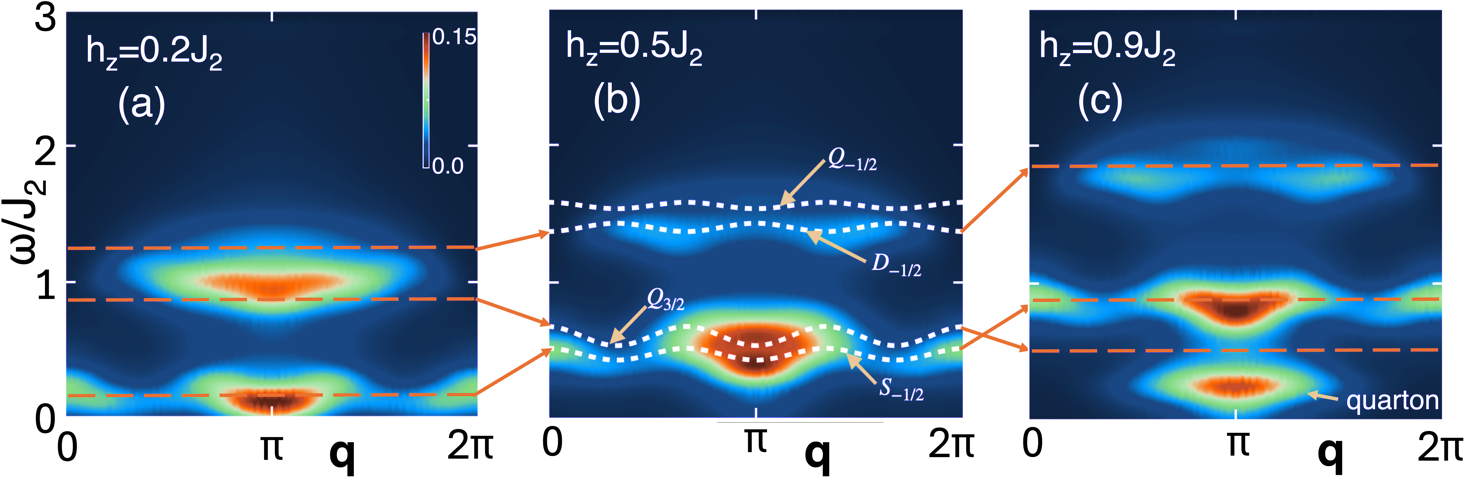}
  \caption{\textbf{RIXS at finite field:} $\chi^0(\mathbf{q},\omega)$ on a 120-site chain for indicated $h_z$ values. Long-dashed lines in (a,c) show analytically predicted feature locations; fine-dashed lines in (b) mark single-excitation dispersions. Orange arrows guide the $h_z$-dependent evolution. Color scale is identical across all panels.}
  \label{fig3}
\end{figure}

At first order ($l=1$), the RIXS operator $O_{i,1}$ reduces to two terms: $J_{i,i\pm1}S^x_{i\pm1}$ and $J_{i,i\pm1}S^z_{i\pm1}S^y_{i\pm1}$ (see \cite{supp}, Sec. IV. C, Eq. 11). The latter, a \textit{non-local} term, enables excitations within a trimer or across neighboring trimers. In the two-trimer system, $O_{i,1}$ can scatter $|SS\rangle$ components of $|g\rangle_{2Tr}$ with $S^z_T = 0$ into $|DD\rangle$, $|DQ\rangle$, and $|QQ\rangle$ states with $\Delta S^z = \pm1$. Consequently, spectral weight from the spinon continua shifts to higher-energy modes, as evident in Fig.~\ref{fig1}(c,d). This redistribution intensifies at $l=2$ [Fig.~\ref{fig1}(e,f)], where the key contribution is from $S(S+1)\frac{5}{4}\sum_{j\ne k} J_{i,j}J_{i,k} S^x_i$, with $j, k$ NN to site $i$ (see \cite{supp}, Sec. IV. C Eq. 12) and $j\neq k$. Notably, Fig.~\ref{fig1}(f) reveals multiplet features at $E_{DQ}$ and $E_{2Q}$, corresponding to $|DQ\rangle$ and $|QQ\rangle$ states (marked by arrows), confirming that higher-order RIXS can directly probe composite excitations.

\textit{RIXS spectra at magnetization plateau:---}  
For the field evolution, we focus on $l=0$ RIXS; higher-order ($l > 0$) terms primarily dress the $l=0$ spectra without introducing new features. We examine $l=0$ RIXS at finite field ($h_z = 0.5J_2$), highlighting field-induced splitting in the single- and two-trimer spectra [Fig.~\ref{fig2}(c,d)]. Compared to Fig.~\ref{fig2}(a), Fig.~\ref{fig2}(c) shows that the single-trimer levels $|S\rangle_\sigma$, $|D\rangle_\sigma$, and $|Q\rangle_\sigma$ split under $h_z$, with spin-aligned (anti-aligned) states shifting down (up), lifting all degeneracies. In the coupled two-trimer case [Fig.~\ref{fig2}(d)], these levels broaden into a quasi-continuum, in contrast to the well-separated bunching levels in Fig.~\ref{fig2}(b).

At $h_z = 0.5J_2$, the two-trimer ground state has total spin $S^z_T = 1$, consistent with the 1/3 magnetization plateau. Allowed $\Delta S^z = \pm 1$ excitations are indicated by blue arrows in Fig.~\ref{fig2}(d), with their energy windows,  relative to the energy ground state, are indicated by horizontal bands. These identify the key spin-flip excitations contributing to the plateau-phase RIXS spectrum despite the quasi-continuous distribution of eigenvalues.

Figures~\ref{fig3}(a–c) show $l=0$ RIXS spectra at $h_z = 0.2J_2$, $0.5J_2$, and $0.9J_2$. We notice that the gap between the ground state and the first excitation evolves non-monotonically, reflecting field-dependent spectral reorganization. We first focus on $h_z = 0.5J_2$, the midpoint of the 1/3 plateau [Fig.~\ref{fig0}(b)]. The RIXS features in Fig.~\ref{fig3}(b) cluster around $\omega \approx 0.5J_2$, $1.5J_2$, and $2.5J_2$, matching the centroids of gray bands in Fig.~\ref{fig2}(d), although the highest feature is very weak. The lowest feature includes $\Delta S^z = -1$ transitions $|S\rangle_{1/2} \to |S\rangle_{-1/2}$ and $\Delta S^z = +1$ transitions $|S\rangle_{1/2} \to |Q\rangle_{3/2}$ (energy $\sim 0.58J_2$). The mid-energy feature at $1.5J_2$ stems from single-trimer transitions to $|D\rangle_{-1/2}$ ($1.37J_2$) and $|Q\rangle_{-1/2}$ ($1.54J_2$). Dashed lines in Fig.~\ref{fig3}(b) trace the dispersion of these single-trimer excitations over the product state $|\tilde{g}\rangle_{\text{plat}} = \prod_{i=1}^{L/3} |S_i\rangle_\uparrow$, confirming their energies (details in \cite{supp} Sec. IV B).
The higher-energy band near $2.5J_2$ in Fig.~\ref{fig2}(d) involves two-trimer excitations, e.g., $|S\rangle_{1/2}|D\rangle_{1/2} \to |D\rangle_{-1/2}|D\rangle_{1/2}$ at $\sim 2.26J_2$, but this feature is barely visible in Fig.~\ref{fig3}(b). This suppression is explained by ED results showing that the plateau ground state is well approximated by the product state $|\tilde{g}\rangle_{\text{plat}}$, leading to weak overlap with two-trimer excited states.

An analytical expression for $\chi^0(\mathbf{q},\omega)$ can be obtained by approximating the trimer ground state $|S_{i}\rangle_{\sigma}$ as $|\sigma,\bar{2,3}\rangle$ in $|\tilde{g} \rangle_{\text{plat}}$, as detailed in~\cite{supp}, Sec. IV.D. This yields:
$\chi^0(\mathbf{q},\omega)=\frac{1}{3}\delta(\omega - h_z) + \frac{2}{3}\sin^2(qa/2)\left\{\delta(\omega - J_2 + h_z/2) + \delta(\omega - J_2 - h_z)\right\}$, 
which captures the three low-energy features in Fig.~\ref{fig3}(b), indicated by the endpoints of the orange arrows on the left edge of the panel. The expression shows that the feature positions evolve linearly with $h_z$. This behavior is confirmed by the dashed lines in Fig.~\ref{fig3}(a) and (c) for $h_z = 0.2J_2$ and $0.9J_2$, respectively. The arrows across panels trace the continuous shift of spectral features with increasing field. 
While the plateau ground state ($\approx |1_\uparrow,\bar{2,3}\rangle$ per trimer) is gapped, the low-energy excitation evolves into $|Q\rangle_{3/2}$ -continuum as seen in Fig.~\ref{fig3}(c). Beyond $h_z = 1.12J_2$, the $|Q\rangle_{3/2}$ crosses over the plateau ground state, forming a novel quarton continuum ground state ~\cite{supp}, Sec. V. For field values slightly above  $h_z = 1.12J_2$, the magnetization exceeds the plateau but remains well below full polarization, suggesting a gradual increase in the population of trimer $|Q\rangle_{3/2}$ states via ($\Delta S=1$) single-trimer excitations involving singlet bond breaking: $|1_\uparrow,\bar{2,3}\rangle \rightarrow |1_\uparrow 2_\uparrow 3_\uparrow\rangle$. This singlet-to-triplet ($S=1$) excitations can undergo condensation akin to triplet condensation in TlCuCl$_3$\cite{bec-1-nat} and KGaCu(PO$_4$)$_2$\cite{bec-3-1d-field}, signaling a field-induced 1D quarton Bose-Einstein condensate (BEC) phase. Further, weak inter-chain coupling in the material can stabilize such condensate \cite{bec-4-2d}, and even for isolated chains, signatures of field-driven such quantum phase transition can survive at finite temperatures\cite{bec-2-rmp}. 

\textit{Conclusion:---}  Our work establishes $\rm{Cu_3(P_2O_6OH)_2}$ as a promising quantum magnet where RIXS directly probes fractionalization and emergent composite excitations in a trimerized spin-1/2 chain. Combining DMRG, ED, and real-space RG, we identify a gapless spinon continuum and a hierarchy of higher-energy quasiparticles—doublons, quartons, and two-trimer composites. The non-local nature of higher-order RIXS processes selectively suppresses spinon weight while enhancing higher-energy quasiparticle signatures, offering direct experimental access to composite excitations.
Within the 1/3 plateau, feature locations shift linearly with the magnetic field. This lowers the energy of spin-polarized quarton excitations, which forms a gapless ground state beyond the plateau. These quartons arise from $\Delta S=1$ transition by exciting singlet states mediated by $J_2$ in a trimer to triplets. Their proliferation with increasing field drives magnetization and potentially enables a field-induced $S=1$ BEC phase. 
Unlike $\rm{Cu_3(P_2O_6OH)_2}$, the $J_1 > J_2$ trimer chain hosts weakly coupled trimers with RIXS spectra spread over a broader energy range (\cite{supp}, Sec. V). This spectral broadening raises the critical field for overcoming the plateau as seen in $\rm{Na_2Cu_3Ge_4O_{12}}$ (~200 T) \cite{cheng-field} compared to 85 T for $\rm{Cu_3(P_2O_6OH)_2}$, making the latter experimentally accessible.

Our work unveils how geometric trimerization and nonlocal probes like RIXS enable generation and access to emergent quasiparticles and field-tunable condensate phases in low-dimensional quantum magnets.

\textit{Acknowledgments:---} We acknowledge the use of NOETHER and KALINGA clusters at NISER. AM acknowledges funding from the Department of Atomic Energy, India under Project No. 12-R\&D-NIS-5.00-0100. SP acknowledges discussions with Prabhakar.

\bibliography{bibfile}{}



\newpage

\clearpage



\newcounter{defcounter}
\setcounter{defcounter}{0}

\setcounter{equation}{0}
\renewcommand{\theequation}{\arabic{equation}}

\setcounter{figure}{0}
\renewcommand{\thefigure}{\arabic{figure}}

\setcounter{page}{1}
\pagenumbering{roman}

\begin{onecolumngrid}
	\begin{center}
		{\fontsize{12}{12}\selectfont
			\textbf{Supplemental Materials: ``Emergent Quasiparticles \& Field-Tuned RIXS Spectra in a Trimerized Spin-1/2 Chain"\\[5mm]}}
		{\normalsize Subhajyoti Pal$^{1,2*}$, Pradeep Thakur$^{1,2}$, Ashis Kumar Nandy$^{1,2}$, and Anamitra Mukherjee$^{1,2}$\\[1mm]}
		{\small  $^1$\textit{School of Physical Sciences, National Institute of Science Education and Research, Jatni 752050, India}\\[1mm]}
		{\small $^2$\textit{Homi Bhabha National Institute, Training School Complex, Anushakti Nagar, Mumbai 400094, India}\\[1mm]}
		{}
	\end{center}
	\normalsize
\end{onecolumngrid}





\begin{onecolumngrid}








\hfill\begin{minipage}{\dimexpr\textwidth-1.5cm}
\parbox{15cm}{This supplement provides supporting details for the main text. We outline the crystallographic structure and exchange-path analysis that motivate the spin-1/2 trimer chain Hamiltonian for $\rm{Cu_3(P_2O_6OH)_2}$. We discuss analytical and variational methods used to construct doublon and quarton dispersions. We discuss the RIXS cross-section up to the second order in the ultra-short core-hole lifetime approximation. We present an approximate analytic form for the $l=0$ RIXS spectrum in the 1/3 plateau to field-dependent evolution. Finally, we also compare $J_1 < J_2$ and $J_1 > J_2$ regimes, highlighting the role of trimer coupling in dictating spectral features.}
\end{minipage}
\vspace{20pt}

\begin{figure}[ht!]
	\centering
	\includegraphics[width=0.5\linewidth]{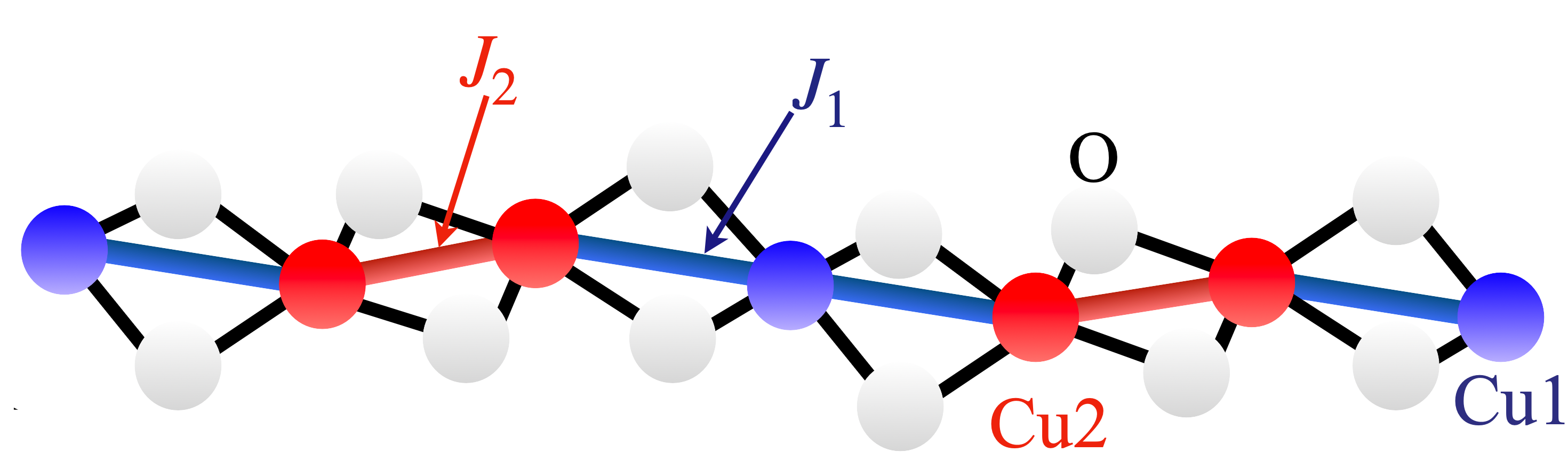}
	\caption{Schematic drawing of positions of Cu and O connecting to Cu in $\rm{Cu_3(P_2O_6OH)_2}$. Blue, red, and white circles, respectively, indicate Cu1 sites, Cu2 sites, and O sites connecting to Cu. Black bars represent Cu-O bonds. Black bars denote Cu-O bonds, while red and blue bars indicate the first and second shortest Cu-Cu bonds, corresponding to the $J_1$ and $J_2$ exchange interactions, respectively.}
	\label{sfig1}
\end{figure}
\section{Hamiltonian for $\rm{Cu_3(P_2O_6OH)_2}$}\label{sec:model}
The crystal structure of $\rm{Cu_3(P_2O_6OH)_2}$ belongs to the space group P1 (No. 2). The lattice parameters are: a = 4.7819(16) \AA, b = 7.0370(8) \AA, c = 8.3574(8) \AA, with angles $\alpha$ = 66.6790(6)\degree, $\beta$ = 76.9930(7)\degree, and $\gamma$ = 72.0642(6)\degree~\cite{hase2007}. The unit cell contains one formula unit (Z = 1). The Cu$^{2+}$ ions have a spin-1/2 in this compound. Supplemental Material Fig: \ref{sfig1} schematically illustrates the positions of Cu and the O atoms that connect to Cu. 
There are two distinct Cu sites, labeled Cu1 and Cu2, and two main Cu-Cu exchange paths. The shortest Cu-Cu distance is 3.06~\AA, featuring two identical Cu-O-Cu superexchange pathways with bond angles of $100.8^\circ$. The second-shortest Cu-Cu distance is 3.28~\AA, with two Cu-O-Cu angles of $98.0^\circ$ and $98.5^\circ$~\cite{hase20061}. Based on the Cu-Cu distances and Cu-O-Cu bond angles, the dominant magnetic interactions are  $J_1$ (first-shortest Cu-Cu bond) and $J_2$ (second-shortest Cu-Cu bond). Other Cu--Cu distances exceed 4.27~\AA, implying weaker interactions. As a result, $\rm{Cu_3(P_2O_6OH)_2}$ can be described as a compound containing spin-1/2 trimer chains with alternating $J_1-J_2-J_1$ interactions. Previous comparison of the experimental magnetization with field with Quantum Monte-Carlo\cite{hase2007} has yielded $J_1=30$K and $J_2=111$K with $J_1/J_2=0.27$. The Hamiltonian has been provided in Eq. 1 of the paper.

\section{RIXS Formalism and the UCL Approximation}

We provide a brief discussion of the theoretical framework for RIXS.  
As mentioned in the paper, in the RIXS process, an incoming X-ray photon excites a core electron (e.g., from a 2$p$ level) into the valence band, creating a short-lived core hole. After a brief evolution, the core hole is filled by a valence electron, emitting an outgoing photon. The energy loss $\omega = \omega_{\mathrm{in}} - \omega_{\mathrm{out}}$ corresponds to an excitation in the material.

The full polarization-dependent RIXS cross-section is described by the Kramers-Heisenberg (KH) formula~\cite{rixs-nat-rev,RevModPhys.83.705}:
\begin{equation}
I_{\epsilon}(\mathbf{q},\omega) =\sum_{F} \left| \langle F |  \, \mathcal{O_{\mathbf{q},\epsilon}} \, | G \rangle \right|^2 \delta(E_F - E_G - \omega),
\end{equation}
where $|G\rangle = |g\rangle \otimes |1\rangle_{\epsilon, \omega_{\mathrm{in}}} \otimes |2p^6\rangle$ and $|F\rangle = |f\rangle \otimes |1\rangle_{\epsilon', \omega_{\mathrm{out}}} \otimes |2p^6\rangle$ denote the initial and final states including valence, photon, and core levels. The energies are
$E_G = E_g + E_{2p^6} + \omega_{\mathrm{in}}$ and $E_F = E_f + E_{2p^6} + \omega_{\mathrm{out}}$.

The operator $\mathcal{O}_{\mathbf{q},\epsilon}=1/\sqrt{L}\sum_je^{i\mathbf{q}.j}\mathcal{O}_{j,\epsilon}$ captures the intermediate-state evolution:
\begin{equation}
\mathcal{O}_{j,\epsilon} = D^\dagger_{j,\epsilon^F}\frac{1}{\omega_{\mathrm{in}} - \mathcal{H} + i\Gamma}D_{j,\epsilon^I}
\end{equation}
where $\mathcal{H} = H + H_c$ includes the valence-band Hamiltonian $H$ and the core-hole potential $H_c$, and $\Gamma$ is the inverse core-hole lifetime. The dipole-allowed transitions are controlled by the initial and final polarization dependent dipole operators $D_{j,\epsilon^I}$ and $D_{j,\epsilon^F}$, where $\epsilon^I$ and  $\epsilon^F$ are the corresponding photon polarizations. 

Since the core-hole state returns to the same configuration as in the final state, it can be integrated out, yielding an effective RIXS operator acting only in the valence sector. Further with the X-ray energy ($\omega_{in}$) in resonance with the Cu $L$-edge  the expression simplifies to ~\cite{PhysRevX.6.021020}:
\begin{equation}
I_{\epsilon}(\mathbf{q},\omega) =|\mathcal{M}_{\epsilon}|^2\sum_f 1/\sqrt{L}\sum_je^{i\mathbf{q}.j}\left| \langle f | \tilde{D}_{\mathrm{in},j} \, \tilde{\mathcal{O}} \, \tilde{D}_{\mathrm{out},j} | g \rangle \right|^2 \delta(E_f - E_g - \omega),
\end{equation}
where $\tilde{\mathcal{O}} = \frac{1}{-H + i\Gamma}$, and $\mathcal{M}_\epsilon$ is the polarization-dependent atomic form factor that accounts for the difference between the incoming and outgoing photon polarizations and $H$ is the \textit{valence band Hamiltonian}. The $|i\rangle$,  $|f\rangle$ , $E_g$ and $E_f$, now refer to the ground state and the excitations of the valence Hamiltonian. 
The $\tilde{D}_{\mathrm{in},j}$ and $\tilde{D}_{\mathrm{out},j}$ are the remnants of the original dipole operator after parts of it are absorbed in defining  $\mathcal{M}_\epsilon$. The form of these are dependent on the scattering process being considered. We refer to literature for a detailed description\cite{PhysRevX.6.021020} which also show that for the NSC channel,  $\tilde{D}_{\mathrm{in},j}$ and $\tilde{D}_{\mathrm{out},j}$ are the identity and the spin flip operator ($S^x_j$) respectively.
Finally, ultra-short core-hole lifetime (UCL) approximation amounts to expanding $\tilde{\mathcal{O}}$ in inverse $\Gamma$, $\tilde{\mathcal{O}}=\sum_{l=0}^\infty\frac{H^l}{(i\Gamma)^{l+1}}$. The UCL approximation is valid for cuprates at the L-edge where $J_2/\Gamma \ll 1$, which holds for the cuprates \cite{UKumar2022, PhysRevX.6.021020, PhysRevB.108.214405}. For generic spin Hamiltonians (of the form $\sum_{i,j}J_{i,j}\mathbf{S}_i.\mathbf{S}_j$, modeling undoped half-filled $d$-level such as the cuprates) and invoking that the RIXS excitation is quasi-local within the fast collision approximation \cite{fast-coll, J.van_den_Brink_2006,van_den_Brink_2007}, the expression of the RIXS intensity can be simplified to: 
\begin{align}
I_\epsilon(\mathbf{q},\omega) &\propto \Bigg[  
\frac{1}{\Gamma^2}\sum_f \left|\left\langle f\left|\frac{1}{\sqrt{L}}\sum_i e^{i\mathbf{q}\cdot \mathbf{R}_{i}}S_i^{x}\right|g\right\rangle\right|^2 \nonumber \\
&\quad + \frac{1}{\Gamma^4}\sum_f \left|\left\langle f\left|\frac{1}{\sqrt{L}}\sum_{i,j} e^{i\mathbf{q}\cdot \mathbf{R}_{i}}J_{i,j}S_i^{x}(\hat{S}_i\cdot  \hat{S}_j) \right|g\right\rangle\right|^2 \nonumber \\
&\quad + \frac{1}{\Gamma^6}\sum_f \left|\left\langle f\left|\frac{1}{\sqrt{L}}\sum_{i,j,k} e^{i\mathbf{q}\cdot \mathbf{R}_{i}}J_{i,j}J_{i,k}S_i^{x}(\hat{S}_i\cdot  \hat{S}_j)(\hat{S}_i\cdot  \hat{S}_k) \right|g\right\rangle\right|^2 + \cdots \Bigg] \nonumber \\
&\quad \times \delta\left(E^f_d - E^g_d - \omega\right) \equiv \sum_{l}\chi^{l}(\mathbf{q},\omega)
\end{align}
Thus for the spin Hamiltonian Eq.1 in the paper Eq. 2 (in the paper) provides the contributions to the RIXS process. We note that we only focus on the many-body aspect of the problem. While atomic form factors are important, we aim here to understand the many-body excitations captured in RIXS.

\section{Scaling analysis and phase characterization}
We characterize the different magnetic phases using the magnetization $M(h_z)$, the scaling of the excitation gap $\delta E_0/L$ with system size, and the decay of static spin-spin correlations $C^x(r)=\langle S^x_{L/2} S^x_{L/2+r} \rangle$ with separation  $r$ between the spins. Main paper Figure~\ref{fig0}(b) shows a $1/3$ magnetization plateau in the range $0.179J_2 \leq h_z \leq 1.1J_2$, and full polarization appears above $h_z \approx 1.3J_2$. Within the plateau and the fully polarized phases, the excitation gap $\delta E_0/L$ extrapolates to a finite value as the system length $L \to \infty$ (e.g. $h_z = 0.5, 1.5$), indicating a gapped phase. In contrast, in the intervals $0 < h_z < 0.12J_2$ and $1.12J_2 < h_z < 1.32J_2$,
the gap closes in the thermodynamic limit (e.g. $h_z = 0.0, 1.2$), indicating a gapless spectrum.
To further support these results, we examine the decay of the correlation function, $C^x(r)$. In the gapped plateau and fully polarized phases, $C^x(r)$ decays exponentially~\cite{koma2007} with $r$ (e.g. $h_z = 0.5$ in the lower-right inset of Fig.~\ref{fig0}(b) of the paper). In the gapless phase (e.g. at $h_z = 0$ in the upper-right inset of Fig.~\ref{fig0}(b)), $C^x(r)$ shows power-law decay, $C^x(r) \sim 1/r^{\eta}$ with $\eta \approx 1$, consistent with a conformal field theory with central charge $c = 1$, characteristic of spin-$\frac{1}{2}$  HAC universality class~\cite{universality-1}. 
We note that the  oscillations in the correlation decays occur due to the periodic arrangement of the couplings $J_1$ (weak) and $J_2$ (strong) in the trimer chain are pronounced in the gapless phase (Top right inset).

\section{Effective Hamiltonian and ED-analysis}

\subsection{Effective Hamiltonian for lowest energy feature spectra:}

In Fig.\ref{fig2}(a) of the paper, we see that the low-energy doublets for a single trimer are a linear combination of a singlet and an unpaired spin, meaning each trimer effectively behaves as a spin-1/2 system. Additionally, there is a clear energy gap separating these doublets from the higher-energy states. Consequently, the low-energy excitations of the trimer chain can be accurately described by an effective Hamiltonian that involves only these states. 
We use the real space renormalization group approach based on a well-known identity~\cite{rg-realspace,rg-delgado} for extracting the effective Hamiltonian:
\begin{equation}
\mathcal{P} \frac{1}{z - {H}} \mathcal{P} = \left( z\mathcal{P} - \mathcal{P} {H} \mathcal{P} - \mathcal{P} {H} \mathcal{Q} \frac{1}{z - \mathcal{Q} {H} \mathcal{Q}} \mathcal{Q} {H} \mathcal{P} \right)^{-1}
\label{placeholder}
\end{equation}

Here, $\mathcal{P} = \prod_i P_i$ is the projection operator onto the low-energy subspace, where $P_i$ projects each trimer onto its doublet ground states $|S\rangle_{\pm 1/2}$. The complementary projection is $\mathcal{Q} = I - \mathcal{P}$, which spans the rest of the Hilbert space. The full Hamiltonian is $H = H_t + H_{tt}$, where $H_t$ contains the intratrimer interactions and $H_{tt}$ the intertrimer couplings. We evaluate the effective Hamiltonian at energy $z = \omega + i\eta$, and extract it specifically at $\omega = 2E_{S}$.

\begin{equation}
    \hat{H}_\text{eff}=J_\text{eff}\sum_{i=1}^{L/3} {\bf \tilde{S}}_i \cdot{\bf \tilde{S}}_{i+1}  +J_\text{eff}'\sum_{i=1}^{L/3} {\bf \tilde{S}}_i \cdot{\bf \tilde{S}}_{i+2},
\end{equation}
${\bf \tilde{S}}$ is the effective spin operator of a single trimer in the reduced Hilbert space. The effective Hamiltonian describes a Heisenberg antiferromagnetic chain (HAC) with effective spin-1/2 degrees of freedom per trimer with NN and NNN spin interactions.  However, $J_\text{eff}=a_0 b_0 (a_0 + b_0)^2 J_1\approx0.16J_1$. The effective NNN exchange has a very small magnitude $J_\text{eff}'  =[a_0 b_0 (a_0 + b_0)^2 \left( b_1 (a_0 + b_0) - a_1 b_0 \right) \left( a_0 (2 a_1 + b_1) + b_0 (a_1 + b_1) \right)]/\Delta E$, where $\Delta E = E_0-E_1$ with $E_0$ being the ground state energy of the uncoupled trimer and $E_1$ being the doublon excitation energy of the  uncoupled trimer.
Numerically, $\sim 10^{-4}J_1$ implying to a very high degree of accuracy that $H_{\text{eff}}$ describes a NN HAC \cite{J1-J2_HAC_ED}. This effective Hamiltonian can successfully describe the gapless low-energy excitations known as spinons.  The low-energy region corresponds to the two-spinon continuum, with the predicted lower boundary given by $\omega_l = \pi J_{eff} |\sin(3\mathbf{q})|/2$ and the upper boundary by $\omega = \pi J_{eff} |\sin(3\mathbf{q}/2)|$. These boundaries are shown in the first Brillouin zone as shown in Fig.~\ref{fig1}(a) in the paper by white dashed lines (and are marked as `spinons').

\subsection{Dispersion relations at zero and finite field}
The excitations of the trimer chain can be classified as spinons, doublons, and quartons. Here, we focus on the doublon and quarton excitations, which correspond to the single trimer excitations, $|D\rangle_\sigma$ and $|Q\rangle_\sigma$. Their dispersion relations are shown in Fig.~\ref{fig1}(a) of the main text (zeroth-order RIXS process), and are constructed using a variational approach.   We assume that the ground state of the spin chain is a product of alternating trimer ground states $\left|S\right\rangle_{\sigma}$ and $\left|S\right\rangle_{\bar{\sigma}}$, such that the full chain ground state is given by $\left| \psi_{\mathrm{G}} \right\rangle = \left| S \right\rangle_{\sigma}^{1} \left| S \right\rangle_{\bar{\sigma}}^{2} \cdots \left| S \right\rangle_{\bar{\sigma}}^{L}$. Here, $\bar{\sigma} = -\sigma$ at zero field, and $\bar{\sigma} = \sigma$ in the $1/3$ magnetization plateau. We focus on excitations above this degenerate manifold and do not apply degenerate perturbation theory. For the doublon or quarton excitation, one trimer is promoted from $\left|S\right\rangle$ to $\left|D\right\rangle$ or $\left|S\right\rangle$ to $\left|Q\right\rangle$ respectively  with $\Delta S^z = \pm 1$. The corresponding excited state is $
\left| {\psi_{D/Q}}\right\rangle_r = \left| S \right\rangle_{\sigma}^{1} \cdots \left| D/Q \right\rangle_{\bar{\sigma}'}^{r} \cdots \left| S \right\rangle_{\bar{\sigma}}^{L},
$  with $|\bar{\sigma}' - \bar{\sigma}| = 1$. We assign momentum via Fourier transform: $\left| \psi_{D}^{q} \right\rangle = \frac{1}{\sqrt{L}} \sum_{r=1}^{L} e^{-iqr} \left| \psi_D \right\rangle_r.$ The complete Hamiltonian, assuming periodic boundary conditions, is composed of the sum of the intra-trimer Hamiltonian $H_m$ and the inter-trimer Hamiltonian $H_{m,m+1}$.
\begin{equation}
H = \sum_{m=1}^{L} H_m + \sum_{m=1}^{L} H_{m,m+1},
\end{equation}
with intra-trimer terms $H_m$ and inter-trimer couplings $H_{m,m+1}$.

The ground state energy is:
\begin{equation}
	\left\langle H \right\rangle_{\rm{G}}=\left\langle \psi_{\rm{G}}\right|H \left| \psi_{\rm{G}}\right\rangle
	= L E_S + \frac{1}{4}a_0 b_0(a_0+b_0)^2L J_1.
\end{equation}
where $a_0$, $b_0$ are amplitudes from the ground state trimer basis.

For the excited state $\left| \psi_{D/Q}^q \right\rangle$, only terms with excitations at the same or neighboring trimers contribute. The energy expectation value becomes:
\begin{eqnarray}
	\left\langle H \right\rangle_{D/Q}=&&\left\langle \psi^{q}_{D/Q}\right|H \left| \psi^{q}_{D/Q}\right\rangle \nonumber \\
	=&& \frac{1}{L} \sum_{r=1}^{L} \sum_{l=1}^{L} \rm{e}^{-\rm{i} q \emph{r}+\rm{i} q \emph{l}}
	\left\langle 0^{2} \cdots 1_{\emph{l}}^{1}  \cdots 0^{2} \right| \sum_{m=1}^{L}H_m +\sum_{m=1}^{L}H_{m,m+1} \left| 0^{2} \cdots 1_{\emph{r}}^{1} \cdots 0^{2} \right\rangle 
\end{eqnarray}
The dispersion relation plotted in the Fig:~\ref{fig1}(a) of the main paper is calculated as $\epsilon_{\rm{q}}^{D/Q}=\left\langle H \right\rangle_{D/Q}-\left\langle H \right\rangle_{G}$.

\subsection{Analysis of two-trimer excitations in zero magnetic field at various orders of RIXS}

To understand the features observed in the RIXS spectra and their dependence on the scattering order, we analyze the two-trimer singlet-like ground state at zero field: $
|\tilde{g}\rangle = \frac{1}{\sqrt{2}}\left(|S\rangle_{-1/2} \otimes |S\rangle_{1/2} - |S\rangle_{1/2} \otimes |S\rangle_{-1/2}\right),
$ where $|S\rangle_{\pm 1/2}$ are single trimer ground states (see Fig.~\ref{fig1}(a) in the main paper).

\textit{Zeroth-order RIXS ($l=0$):} For $l=0$, a weak feature appears at $\omega = 1.8J_2$ in main paper Fig.~\ref{fig1}(a,b), matching $2E_D$, the energy of two doublon excitations. However, since the zeroth-order RIXS operator involves only on-site spin flips, it cannot directly create two doublons from $|\tilde{g}\rangle$. Such transitions are only possible through small admixtures of $|SD\rangle$ in the ground state, which are suppressed by the $|E_{2S} - E_{2D}|$ gap. Therefore, the $|DD\rangle$ signal is weak.

\textit{First-order RIXS ($l=1$):} The first-order operator $O_{l=1} = \sum_i O^i_{l=1} e^{i \mathbf{q} \cdot \mathbf{r}_i}$ (see below Eq.~\ref{eq:RIXS_Int_f} of the main text) contains terms involving both single-site and nearest-neighbor spin combinations:
\begin{align}
O_{l=1} &= \frac{1}{4} \sum_i \left( J_{i,i-1} S^x_{i-1} + J_{i,i+1} S^x_{i+1} \right)e^{i \mathbf{q} \cdot \mathbf{r}_i} + i \sum_i \left[ J_{i,i-1} S^z_i S^y_{i-1} (1 - e^{-i\mathbf{q}\cdot\mathbf{a}}) + J_{i,i+1} S^z_i S^y_{i+1} (1 - e^{i\mathbf{q}\cdot\mathbf{a}}) \right] e^{i \mathbf{q} \cdot \mathbf{r}_i}. 
\end{align}
These terms can act either on a single trimer or both simultaneously. As in the $l=0$ case, single-trimer actions yield negligible $|DD\rangle$ weight. But for two-trimer operations, transitions to $|DD\rangle$, $|DQ\rangle$, and $|QQ\rangle$ are allowed, although their matrix elements remain small. The most significant contributions are to $|D\rangle_{-1/2}|D\rangle_{-1/2}$ and $|D\rangle_{1/2}|D\rangle_{1/2}$.

\textit{Second-order RIXS ($l=2$):} The second-order operator simplifies to:
\begin{align}
O_{l=2} = \sum_{j,k} J_{i,j}J_{i,k} S^x_i \left(\mathbf{S}_i \cdot \mathbf{S}_j\right) \left(\mathbf{S}_i \cdot \mathbf{S}_k\right) 
= \frac{5}{4} \sum_{j,k} J_{i,j} J_{i,k} S^x_i (\mathbf{S}_j \cdot \mathbf{S}_k) 
- i \sum_{j,k} J_{i,j} J_{i,k} S^x_i \mathbf{S}_i \cdot (\mathbf{S}_j \times \mathbf{S}_k). 
\end{align}
The second term (scalar chirality) vanishes in non-chiral systems. For the first term, only the $j \neq k$ contributions lead to nontrivial inter-trimer excitations. 

\subsection{RIXS ($l=0$) spectra at magnetization plateau}

A simplified understanding of the $q$-dependence in the low-energy $l=0$ RIXS spectra can be obtained by approximating $|S\rangle_{1/2} \approx |\uparrow, \bar{2,3}\rangle$ in $|\tilde{g}\rangle_{plat}$, we note that this is  reasonable approximation as $a_0\gg b_0$ in $|S\rangle_{1/2}(=a_0|\uparrow\bar{23}\rangle+a_0|\bar{12}\uparrow\rangle)$, as seen from Fig 3 (a) in the main paper where the values are provided in the caption. In the magnetic field ($h_z=0.5J_2$) this state has an energy $E_{\tilde{g}}=-h_z/2(=-0.25J_2)$.

Allowed excitations (with $\Delta S^z = \pm 1$) are $\{|f\rangle=|\downarrow\bar{23}\rangle,|\uparrow\uparrow\uparrow\rangle,|\uparrow\downarrow\downarrow\rangle\}$ with energies (of $h_z$, $J_2 - h_z/2$, and $J_2 + h_z$, respectively, measured from $E_{\tilde{g}}$ . The resulting analytical  $l=0$ RIXS spectra admits a simple form, $\frac{1}{3}\delta(\omega-h_z)+\frac{2}{3} \sin^2(qa/2)\{\delta(\omega-J_2+h_z/2)+\delta(\omega-J_2-h_z)\}$. Hence, in Fig.~\ref{fig3}(b), the non-dispersive feature at $\omega = h_z$ corresponds to the first term, while the dispersive peaks at $\omega = 0.75J_2$ and $1.5J_2$ near $q = \pi$ arise from the remaining terms. \textit{We note that only the lower three bands are captured within this simple approach.}
The $h_z$ dependence of the eigenvalues of these final states for the three field values in Fig.~\ref{fig3} are as follows:

\begin{table}[ht!]
\centering
\begin{tabular}{|c|c|c|c|c|}
\hline
Final State & $h_z = 0.2J_2$ & $h_z = 0.5J_2$ & $h_z = 0.9J_2$& State Description \\
\hline
$h_z$ & 0.2 & 0.5 & 0.9& $|\downarrow \bar{23} \rangle \approx |S\rangle_{-1/2}$ \\
\hline
$J_2 - h_z/2$ & 0.9 & 0.75 & 0.55& $|\uparrow \uparrow \uparrow \rangle = |Q\rangle_{3/2}$ \\
\hline
$J_2 + h_z$ & 1.2 & 1.5 & 1.9& $|\uparrow \downarrow \downarrow \rangle \approx |D\rangle_{-1/2}$ \\
\hline
\end{tabular}
\caption{Energies of single trimer excitations for different $h_z$.}
\end{table}

\section{Comparing with RIXS for $J2<J_1$}
\begin{figure}[t]
	\centering
	\includegraphics[width=0.6\linewidth]{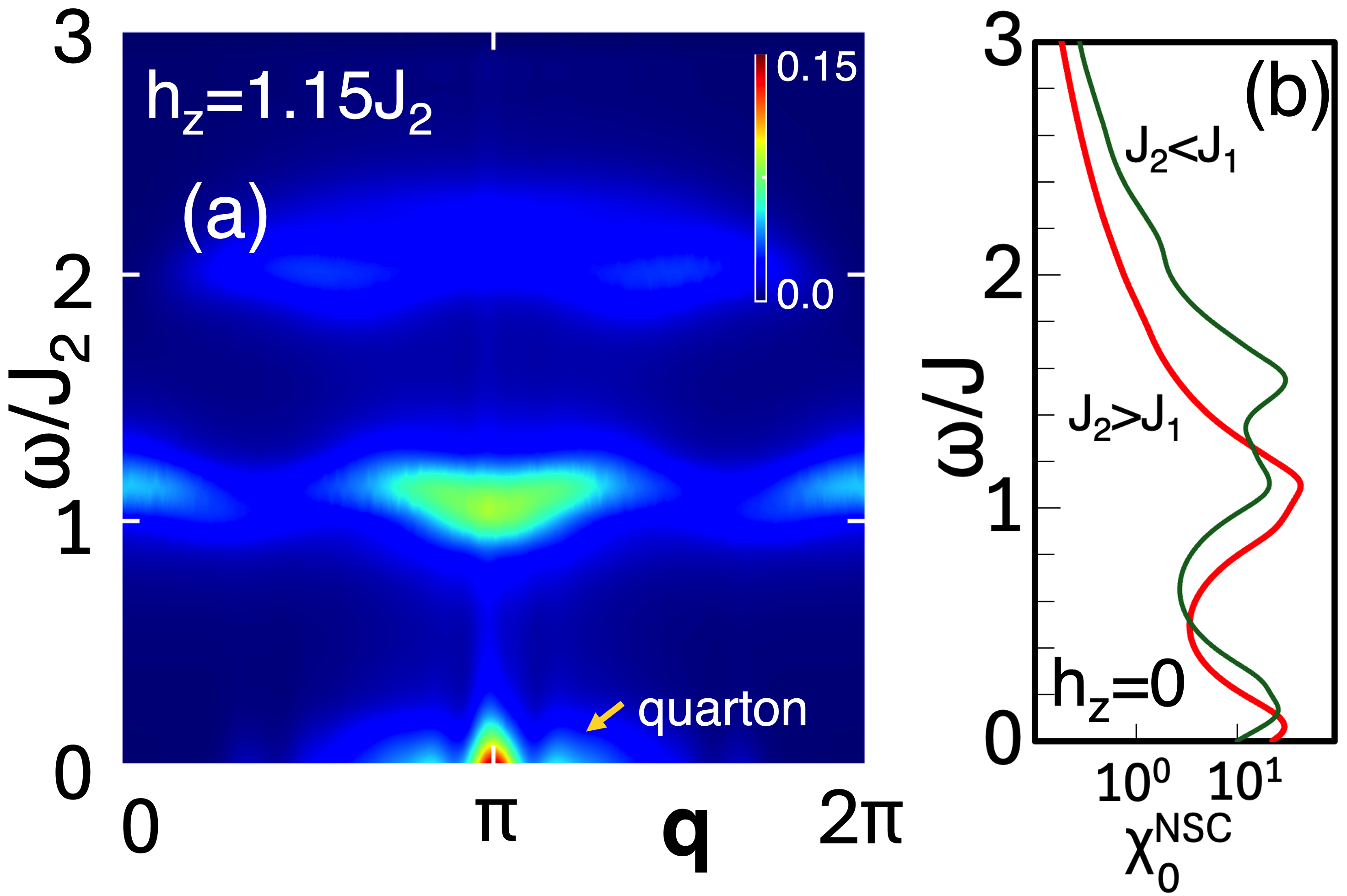}
	\caption{(a) $\chi^0(\mathbf{q},\omega)$ for $h_z=1.15J_2$, beyond the plateau with a gapless spectra ans quarton $|Q\rangle_{3/2}$ continuum. (b) Zero field $\chi^0(\omega)$ for trimerized spin-1/2 chain with $J_1>J_2$ ($J_2/J_1=0.27$) and $J_1<J_2$ ($J_1/J_2=0.27$).}
	\label{sfig2}
\end{figure}
Supplemental Materials Fig.~\ref{sfig2} (a) shows $\chi^0(\mathbf{q},\omega)$ for $h_z=1.15J_2$ a field value beyond the plateau, where the spectrum is gapless with $|Q\rangle_{3/2}$ continuum.
In Supplemental Materials Fig.~\ref{sfig2} (b), we compare the $l=0$ contribution to the RIXS spectra presented in the paper with that for $J_2>J_1$. We see that the weakly coupled trimers have a greater spread of the single trimer spectra, which results in larger spectral gaps and quite different spectral features. These are discussed in literature\cite{cheng2022fractional,prabhakar2024}. Consequently, the magnetic field-induced plateau occurs at greater field values\cite{cheng-field}.

\end{onecolumngrid}

\end{document}